# Distribution asset management through coordinated microgrid scheduling

Mohsen Mahoor[1] ✉, Alireza Majzoobi[1], Amin Khodaei[1]
[1]Department of Electrical and Computer Engineering, University of Denver, 2155 E. Wesley Ave, Denver, CO 80210, USA
✉ E-mail: mohsen.mahoor@du.edu

**Abstract:** Distribution Asset Management is an important task performed by utility companies to prolong the lifetime of the critical distribution assets and to accordingly ensure grid reliability by preventing unplanned outages. This study focuses on microgrid applications for distribution asset management as a viable and less expensive alternative to traditional utility practices in this area. A microgrid is as an emerging distribution technology that encompasses a variety of distribution technologies including distributed generation, demand response, and energy storage. Moreover, the substation transformer, as the most critical component in a distribution grid, is selected as the component of the choice for asset management studies. The resulting model is a microgrid-based distribution transformer asset management model in which microgrid exchanged power with the utility grid is reshaped in such a way that the distribution transformer lifetime is maximised. Numerical simulations on a test utility-owned microgrid demonstrate the effectiveness of the proposed model to reshape the loading of the distribution transformer at the point of interconnection in order to increase its lifetime.

## Nomenclature

### Indices

| | |
|---|---|
| $d$ | index for loads |
| $h$ | index for day |
| $i$ | index for DERs |
| $t$ | index for hour |

### Superscripts

| | |
|---|---|
| ch | distributed energy storage (DES) charge |
| dch | DES discharge |
| H | winding hottest-spot |
| I | initial value of variables and parameters |
| R | rated value |
| TO | transformer top oil |
| U | ultimate value of variables and parameters |
| W | transformer winding |
| ^ | superscript for calculated/given variables |

### Sets

| | |
|---|---|
| $D$ | set of adjustable loads |
| $G$ | set of dispatchable units |
| $S$ | set of energy storage units |

### Parameters

| | |
|---|---|
| DR | ramp down rate |
| DT | minimum down time |
| $E$ | load total required energy |
| $F(.)$ | generation cost |
| $F_{AA}$ | aging acceleration factor of insulation |
| $F_{AA,n}$ | aging acceleration factor for the temperature which exists during the time interval $\Delta t_n$ |
| $F_{EQA}$ | equivalent aging factor for the total time period |
| $M$ | large positive constant |
| MC | minimum charging time |
| MD | minimum discharging time |
| MU | minimum operating time |
| $m/n$ | empirically derived exponent to calculate the variation of $\Delta\theta^H/\Delta\theta^{TO}$ with changes in load |
| $R$ | ratio of full-load loss to no-load loss |
| $\Delta t$ | time interval |
| UR | ramp up rate |
| UT | minimum up time |
| $w$ | binary islanding indicator (1 if grid-connected, 0 if islanded) |
| $\alpha, \beta$ | specified start and end times of adjustable loads |
| $\rho$ | market price |
| $\eta$ | energy storage efficiency |
| $\psi$ | transformer investment cost |
| $\tau$ | time period |
| $\theta$ | temperature (°C) |

### Variables

| | |
|---|---|
| $C$ | energy storage available (stored) energy |
| $D$ | load demand |
| $I$ | commitment state of dispatchable units |
| $K$ | transformer loading ratio |
| $P$ | DER output power |
| $P^M$ | utility grid power exchange with the microgrid |
| $P^{M1}, P^{M2}$ | slack variable for utility grid power |
| $Q$ | cost of transformer loss of life |
| $T^{ch}$ | number of successive charging hours |
| $T^{dch}$ | number of successive discharging hours |
| $T^{on}$ | number of successive ON hours |
| $T^{off}$ | number of successive OFF hours |
| $u$ | energy storage discharging state (1 when discharging, 0 otherwise) |
| $v$ | energy storage charging state (1 when charging, 0 otherwise) |
| $x, y$ | binary variables for selecting slack variables associated with utility grid exchange |
| $z$ | adjustable load state (1 when operating, 0 otherwise) |
| $\lambda, \mu$ | dual variables |
| $\Lambda$ | reflected cost for transformer loss of life in the master problem |

## 1 Introduction

Asset management denotes management and engineering practices applied to valuable assets of a system in order to deliver the required level of service to the customers. Asset management has always been a critical responsibility of electric utility companies to maintain network reliability and quality of service at acceptable levels by reducing the failure probability of critical grid components. In other words, asset management extends the lifetime of equipment and decreases the risk of equipment failure and unplanned power outages. Considering that the current power grid is mainly built in the 1950s and 60s and at the same time the customers' expectations of a high quality of service are at all-time high, the topic of asset management has become more important than ever [1–4].

Transformers are one of the most important electrical equipment when it comes to asset management, conceivably due to their impact on power system adequacy and reliability. Transformer failures can potentially lead to unplanned power outages, in addition to costly and time-consuming repair and replacement [2–4]. Condition monitoring, online monitoring, routine diagnostic, scheduled maintenance, and condition based maintenance (CBM) are some of the most common transformer asset management methods [2, 5, 6]. The lifetime of a transformer highly depends on its insulation condition owing to a higher probability of insulation failure compared with its other components. Moreover, aging of transformer insulation is a function of insulation moisture, oxygen amount, and internal temperature specifically at the hottest spot, which is mainly governed by transformer loading and ambient temperature [7–9]. In [4], power transformer asset management is performed using a two-stage maintenance scheduler. The effect of temperature, thermal aging factors, and electrical aging factors on transformer insulation is experimentally analysed in [10]. In [11], an experimental thermal model for 25 kVA transformers is proposed which estimates transformer lifetime and accordingly the time of transformer maintenance or replacement.

A method for calculating transformer insulation loss of life is provided as a standard, IEEE Std. C57.91-2011 Guide for Loading Mineral-Oil-Immersed Transformers, in [12]. The authors in [13] present a sensory model framework in which transformer lifetime is estimated based on the measured values of winding hottest-spot temperature and the aforementioned IEEE standard. The study in [7] proposes a model for estimating the remaining life of transformer insulation via this IEEE standard, based on historical data of load and ambient temperature. A fuzzy modelling in [14] is applied for transformer asset management while improvement in the remaining life of a transformer is achieved by a fuzzy model system. Application of different machine learning methods, such as Adaptive Network-Based Fuzzy Inference System (ANFIS), Multi-Layer Perceptron (MLP) network and Radial Basis Function (RBF) network, in estimating transformer loss of life is presented in [15], where further these methods are fused together to improve the estimation accuracy [16]. In [17], an artificial neural network is modelled to predict top oil temperature in a transformer, where ambient temperature and load current are considered as the input layer and top oil temperature as the output layer. Since transformer loading has the most significant effect on transformer insulation loss of life, its management and control can remarkably increase transformer lifetime. In [8, 18, 19], the effect of electric vehicles on distribution net load profile and accordingly on distribution equipment such as transformers is studied, and a smart charging method is proposed to manage distribution and transmission assets, including transformers, via controlling and managing distribution net load profile. The effect of electric vehicles and rooftop solar photovoltaic on distribution transformer aging is investigated in [20, 21]. These studies show that rooftop solar generation decreases transformer loss of life, as it reduces the power transferred from the utility grid to loads, while electric vehicles increase transformer loss of life and their charging/discharging should be controlled to prevent negative impacts on the connected transformer's lifetime. A control algorithm with the objective of controlling the electric load of the plug-in electric vehicle on a distribution transformer is proposed in [22]. The proposed algorithm aims at reducing distribution transformer overloading via leveraging vehicle-to-gird strategy. An electric vehicle charging algorithm is studied in [23] in order to coordinate the gird and distribution transformer. The algorithm is able to prevent the distribution transformer from overloading and sharp ramping through smoothing the transformer load profile. In this paper, a new method for distribution transformer asset management by leveraging microgrids is proposed. The microgrid, as defined by the U.S. Department of Energy, is 'a group of interconnected loads and Distributed Energy Resources (DERs) within clearly defined electrical boundaries that acts as a single controllable entity with respect to the grid and can connect and disconnect from the grid to enable it to operate in both grid-connected or island-mode' [24]. Microgrids provide both consumers and utility companies with significant advantages including, but not limited to, improved resiliency and reliability, reduced emission, improved power quality, and enhanced energy efficiency. Microgrids can be operated in either islanded or grid-connected mode. A microgrid in the default operation mode, i.e., grid-connected, is able to exchange power with the utility grid based on its economic objectives [25–27]. In case of faults and/or disturbances in the upstream network, islanded mode plays an active role in microgrid operation, where the microgrid can be intentionally disconnected from the utility grid in order to face the minimum load curtailment [28–31]. In recent years, microgrid deployment has been meaningfully increased and it can be expected that the growing trend is even becoming faster in the near future [32, 33], expected to reach a global revenue of $19.9 billion by 2020 [34]. This trend advocates on the growing interest in microgrids as a mainstay of future power grids. A comprehensive survey on microgrid research trends can be found in [35]. This paper builds upon existing research and deployment efforts and focuses on the flexibility advantages of the utility-owned microgrids as a complementary value proposition for distribution transformer asset management. The microgrid capability in managing its adjustable loads, dispatchable distributed generation (DG) units, distributed energy storage (DES) units, and the ability of exchanging power with the utility grid in the grid-connected mode is specifically considered in this paper for smoothing distribution transformer loading, and consequently decreasing transformer loss of life which leads to higher transformer lifetime. It is assumed that the studied microgrid is utility-owned, thus can be scheduled by the electric utility company or any designated entity as the operator.

### 1.1 Paper contribution

By leveraging the IEEE Std. C57.91-2011, the distribution transformer loss of life is calculated in order to be integrated into the microgrid optimal scheduling model. The aforementioned standard for calculation of the distribution transformer loss of life has a set of nonlinear equations which would make the microgrid optimal scheduling a non-linear and hard to solve the problem. To ensure that the microgrid optimal scheduling problem keeps its linear characteristics, the original problem is decomposed into a mixed integer linear programming master problem (minimising the microgrid operation cost) and a non-linear subproblem (determines the distribution transformer loss of life) using Benders decomposition. These two problems are further coordinated through Benders cuts in an iterative manner. Using this proposed iterative method, the master problem solves the microgrid optimal scheduling problem, as discussed in many existing research such as [28–30], while the added subproblem acts as feedback on how microgrid operation would impact the transformer lifetime, and accordingly, would provide a signal (the Benders cut) on how microgrid schedule should change to increase transformer lifetime. It should be noted that although the proposed models are based on the IEEE Std. C57.91-2011, any other standard or updates to this standard can be modelled using the same approach and without loss of generality in the proposed model.

The remaining of the paper is organised as follows. Section 2 introduces the IEEE Standard for transformer loss of life calculation. The distribution transformer asset management model outline and the problem formulation, including microgrid optimal scheduling master problem and distribution transformer asset management subproblem, are developed in Section 3. The effectiveness of the proposed model is investigated in Section 4 through numerical simulations on a sample microgrid. Finally, the paper is concluded in Section 5.

## 2 IEEE standard – a guide for loading mineral-oil-immersed transformers

The IEEE Std.C57.91-2011 proposes a set of non-linear functions to calculate the transformer loss of life. Equation (1) formulates aging acceleration factor ($F_{AA}$) for a given load and ambient temperature, where $\theta^H$ is a function of transformer load profile and ambient temperature

$$F_{AA} = \exp\left(\frac{15000}{383} - \frac{15000}{\theta^H + 273}\right). \quad (1)$$

Equation (1) is used in calculating the equivalent aging of the transformer in the desired time interval as in (2), which can be considered as one unit of time (can be hour, day, month etc.)

$$F_{EQA} = \frac{\sum_{n=1}^{N} F_{AA_n} \Delta t_n}{\sum_{n=1}^{N} \Delta t_n}, \quad (2)$$

where $\Delta t_n$ is a time interval, $n$ is the time interval index and $N$ is the total number of time intervals. Accordingly, the percentage of insulation loss of life (LOL) is calculated as follows:

$$\text{LOL}(\%) = \frac{F_{EQA} \times t \times 100}{\text{Normal insulation life}}. \quad (3)$$

Based on the IEEE Std. C57.91-2011 [12], normal lifetime for insulation of distribution transformers is 180,000 h. As it can be seen in (1), $\theta^H$ is the backbone term to calculate transformer loss of life. Based on (4), the hottest-spot temperature is composed of three distinct terms

$$\theta^H = \theta^A + \Delta\theta^{TO} + \Delta\theta^H, \quad (4)$$

where $\theta^A$ represents the ambient temperature, $\Delta\theta^{TO}$ is the top-oil rise over ambient temperature, and $\Delta\theta^H$ is the winding hottest-spot rise over top-oil temperature. $\Delta\theta^{TO}$ and $\Delta\theta^H$ are defined in (5) and (6), respectively,

$$\Delta\theta^{TO} = (\Delta\theta^{TO,U} - \Delta\theta^{TO,I})\left(1 - \exp\left(-\frac{1}{\tau^{TO}}\right)\right) + \Delta\theta^{TO,I} \quad (5)$$

$$\Delta\theta^H = (\Delta\theta^{H,U} - \Delta\theta^{H,I})\left(1 - \exp\left(-\frac{1}{\tau^W}\right)\right) + \Delta\theta^{H,I} \quad (6)$$

Moreover, (7)–(10) calculate the initial and the ultimate values for $\Delta\theta^{TO}$ and $\Delta\theta^H$:

$$\Delta\theta^{TO,I} = \Delta\theta^{TO,R}\left[\frac{(K^I)^2 R + 1}{R + 1}\right]^n, \quad (7)$$

$$\Delta\theta^{TO,U} = \Delta\theta^{TO,R}\left[\frac{(K^U)^2 R + 1}{R + 1}\right]^n, \quad (8)$$

$$\Delta\theta^{H,I} = \Delta\theta^{H,R}(K^I)^{2m}, \quad (9)$$

$$\Delta\theta^{H,U} = \Delta\theta^{H,R}(K^U)^{2m}, \quad (10)$$

$K^I$ and $K^U$ are, respectively, the initial and ultimate values of transformer load ratio at each time interval. Both $m$ and $n$ can vary between 0.8 and 1 based on the transformer cooling mode [12, Table 4].

Taking aforementioned equations into account, it can be seen that the percentage value for loss of life at each time interval is a non-linear function of initial/ultimate values of transformer load ratio, and ambient temperature, i.e. $K^I$, $K^U$ and $\theta^A$, respectively. In other words, by knowing $K^I$, $K^U$ and $\theta^A$ at each time interval, the percentage value for loss of life can be calculated via the sequence of these non-linear functions. One key point is that $\theta^A$ can be forecasted accurately for each location at each time interval so that the percentage value for loss of life, as defined in (11), will be a non-linear function of initial and ultimate values of transformer load ratio, i.e., $K^I$ and $K^U$, respectively,

$$\text{LOL}(\%) = f(K_{ht}^I, K_{ht}^U) \quad \forall h, \forall t, \quad (11)$$

## 3 Transformer asset management via microgrid optimal scheduling

The proposed extended microgrid optimal scheduling problem determines the least-cost schedule of available resources (DERs and loads) while minimising the cost of distribution transformer loss of life (12), subject to prevailing operational constraints (13)–(39)

$$\min \sum_h \sum_t \left[\sum_{i \in G} F_i(P_{iht}) + \rho_{ht}^M P_{ht}^M\right] + \psi f(K_{ht}^I, K_{ht}^U) \quad (12)$$

$$\sum_i P_{iht} + P_{ht}^M = \sum_d D_{dht} \quad \forall h, \forall t, \quad (13)$$

$$-P^{M,\max} w_{ht} \leq P_{ht}^M \leq P^{M,\max} w_{ht} \quad \forall h, \forall t, \quad (14)$$

$$P_i^{\min} I_{iht} \leq P_{iht} \leq P_i^{\max} I_{iht} \quad \forall i \in G, \forall h, \forall t, \quad (15)$$

$$P_{iht} - P_{ih(t-1)} \leq UR_i \quad \forall i \in G, \forall h, \forall t \neq 1, \quad (16)$$

$$P_{ih1} - P_{i(h-1)T} \leq UR_i \quad \forall i \in G, \forall h, \forall t, \quad (17)$$

$$P_{ih(t-1)} - P_{iht} \leq DR_i \quad \forall i \in G, \forall h, \forall t \neq 1, \quad (18)$$

$$P_{i(h-1)T} - P_{ih1} \leq DR_i \quad \forall i \in G, \forall h, \forall t, \quad (19)$$

$$T_i^{on} \geq UT_i(I_{iht} - I_{ih(t-1)}) \quad \forall i \in G, \forall h, \forall t \neq 1, \quad (20)$$

$$T_i^{on} \geq UT_i(I_{ih1} - I_{i(h-1)T}) \quad \forall i \in G, \forall h, \forall t, \quad (21)$$

$$T_i^{off} \geq DT_i(I_{ih(t-1)} - I_{iht}) \quad \forall i \in G, \forall h, \forall t \neq 1, \quad (22)$$

$$T_i^{off} \geq DT_i(I_{i(h-1)T} - I_{ih1}) \quad \forall i \in G, \forall h, \forall t, \quad (23)$$

$$P_{iht} \leq P_{iht}^{dch,\max} u_{iht} - P_{iht}^{ch,\min} v_{iht} \quad \forall i \in S, \forall h, \forall t, \quad (24)$$

$$P_{iht} \geq P_{iht}^{dch,\min} u_{iht} - P_{iht}^{ch,\max} v_{iht} \quad \forall i \in S, \forall h, \forall t, \quad (25)$$

$$u_{iht} + v_{iht} \leq 1 \quad \forall i \in S, \forall h, \forall t, \quad (26)$$

$$C_{iht} = C_{ih(t-1)} - (P_{ith} u_{iht} \tau^{ES}/\eta_i) - P_{iht} v_{iht} \tau^{ES} \quad \forall i \in S, \forall h, \forall t \neq 1, \quad (27)$$

$$C_{ih1} = C_{i(h-1)T} - (P_{ih1} u_{iht} \tau^{ES}/\eta_i) - P_{ih1} v_{iht} \tau^{ES} \quad \forall i \in S, \forall h, \forall t, \quad (28)$$

$$C_i^{\min} \leq C_{iht} \leq C_i^{\max} \quad \forall i \in S, \forall h, \forall t, \quad (29)$$

$$T_{iht}^{ch} \geq MC_i(u_{iht} - u_{ih(t-1)}) \quad \forall i \in S, \forall h, \forall t \neq 1, \quad (30)$$

$$T_{ih1}^{ch} \geq MC_i(u_{ih1} - u_{i(h-1)T}) \quad \forall i \in S, \forall h, \forall t, \quad (31)$$

$$T_{iht}^{dch} \geq MD_i(v_{iht} - v_{ih(t-1)}) \quad \forall i \in S, \forall h, \forall t \neq 1, \quad (32)$$

$$T_{ih1}^{dch} \geq MD_i(v_{ih1} - v_{i(h-1)T}) \quad \forall i \in S, \forall h, \forall t, \tag{33}$$

$$D_d^{min} z_{dht} \leq D_{dht} \leq D_d^{max} z_{dht} \quad \forall d \in D, \forall h, \forall t, \tag{34}$$

$$T_d^{on} \geq MU_d(z_{dht} - z_{dh(t-1)}) \quad \forall d \in D, \forall h, t \neq 1, \tag{35}$$

$$T_d^{on} \geq MU_d(z_{dh1} - z_{d(h-1)T}) \quad \forall d \in D, \forall h, \forall t, \tag{36}$$

$$\sum_{[\alpha, \beta]} D_{dht} = E_d \quad \forall d \in D, \tag{37}$$

$$(|\hat{P}_{ht}^M|/P_{nom}^{Trans}) = K_{ht}^U \quad \forall h, \forall t, \tag{38}$$

$$(|\hat{P}_{h(t-1)}^M|/P_{nom}^{Trans}) = K_{ht}^I \quad \forall h, \forall t. \tag{39}$$

The first term in the objective function (12) minimises the microgrid annual operation cost, including the local generation cost and the cost of energy exchange with the utility grid. The second term represents the cost of distribution transformer loss of life. This term consists of a multiplication of distribution transformer loss of life, based on the IEEE Standard explained in Section 2, and the distribution transformer investment cost ($\psi$). This term attempts to minimise the distribution transformer loading in order to reduce its loss of life and consequently increase its lifetime. This investment cost is used to ensure that both terms in the objective have a similar unit (here $). It should also be noted that the maintenance cost of generation units has been already included in the first term of the objective function (12) as the local generation cost.

The load balance (13) ensures that the summation of power exchange with the utility grid and the local generations (including dispatchable DGs, non-dispatchable DGs, and the DES) would be equal to microgrid total load at each operating hour. The DES power can be positive (discharging), negative (charging) or zero (idle). In addition, the power exchange between the microgrid and the utility grid ($P^M$) could be positive (import), negative (export) or zero. This power is also restricted to the capacity of the line between the microgrid and the utility grid (14). Hourly generation of dispatchable DGs is constrained by the maximum and minimum capacity limits (15), where the unit commitment state variable $I$ would be 1 when the unit is committed and 0 otherwise. Constraints (16)–(19) represent ramp up and ramp down constraints of dispatchable DG units, where (16) and (18) belong to intra-day intervals and (17) and (19) represent ramping constraints for inter-day intervals. Dispatchable DG units are subject to the minimum up and down time limits, represented by (20)–(23). Constraints (20), (22) and (21), (23) represent the minimum up/down time for inter-day and intra-day intervals, respectively. Constraints (24) and (25), respectively, define the minimum and maximum limits of the DES charging and discharging. It should be noted that in the charging/discharging mode the binary charging/discharging state variable $v/u$ is 1/0 and the binary discharging/charging state variable $u/v$ is 0/1. Constraint (26) ensures that the DES can merely operate in one mode of charging or discharging at every time period. The amount of charged and discharged power in the DES and the available stored energy determine the stored energy in intra-day (27) and inter-day (28) intervals, where one hour is considered for a time period of charging and discharging. The amount of stored energy in DES is further limited to its capacity (29). Constraints (30), (32) and (31), (33) represent the minimum charging/discharging times of DES for intra-day and inter-day intervals, respectively. Constraint (34) confines adjustable loads to minimum and maximum rated powers, and (35), (36) represent the minimum operating time of adjustable loads for intra-day and inter-day intervals. It should be noted that in (34)–(36), when the load is on, binary operating variable $z$ is 1, otherwise, it is 0. Moreover, (37) considers the required energy to complete an operating cycle for adjustable loads. Note that the adjustable loads utilised in this paper are responsive to price changes and controlling signals from the microgrid controller so that no compensation costs are considered. It should be mentioned that $b =$ 0, which would appear in (17), (19), (21), (23), (28), (31), (33), (36), represents the last day of the previous scheduling horizon, and $T$ represents the last scheduling hour, i.e. $T = 24$ h.

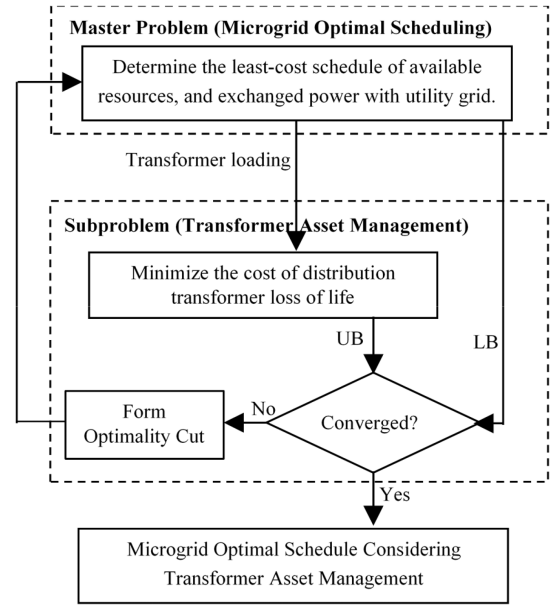

**Fig. 1** *Proposed flowchart for microgrid-based transformer asset management*

As the exchanged power between the microgrid and the utility grid ($P^M$) determines the distribution transformer load ratio, i.e. $K^U$ and $K^I$, constraints (38) and (39) are developed to show the interdependency of these variables. Based on the direction of power exchange between the microgrid and the utility grid, the amount of $P^M$ could be positive (exporting power) or negative (importing power), but the transformer load ratio ($K^I$ or $K^U$) accepts just positive values. Thus, the absolute value of $P^M$ should be considered in (38) and (39), which represent the relationship between the transformer loading and the microgrid power exchange with the utility grid.

### 3.1 Transformer asset management model outline

Fig. 1 depicts the flowchart of the proposed microgrid-based distribution transformer asset management model by using Benders decomposition. The objective of the original microgrid-based distribution transformer asset management model is the summation of microgrid operation cost and the distribution transformer cost of loss of life, i.e. the summation of a linear and a non-linear term. However, in Benders decomposition, the subproblem does not need to be necessarily in a linear form [36]. In this paper, Benders decomposition is employed to decompose the microgrid-based distribution transformer asset management problem to a mixed integer linear programming master problem (minimising the microgrid operation cost) and a nonlinear subproblem (determines the distribution transformer loss of life). These two problems are further coordinated through optimality cuts in an iterative manner. Using this proposed iterative method, the master problem solves the microgrid optimal scheduling problem, while the added subproblem acts as a feedback on how microgrid operation would impact the transformer lifetime, and accordingly, would provide a signal (the optimality cut) on how microgrid schedule should change to increase transformer lifetime. The procedure for microgrid-based distribution transformer asset management solution is as follows:

(i) Solve the microgrid optimal scheduling master problem by considering the commitment and dispatch of available DGs, the charging and discharging schedules of DESs, the schedule of adjustable loads, and the exchanged power with the utility grid. Note that there is no optimality cut available in the first iteration of the master problem.

(ii) Minimise the transformer asset management subproblem by considering the exchanged power of the microgrid with the utility grid (transformer loading).

(iii) Compare the subproblem's solution, i.e. an upper bound, with the solution of the master problem, i.e. a lower bound. If the difference is larger than a predetermined threshold, form the optimality cut and send back to the master problem to consequently revise the current schedule of available resources and the exchanged power with the utility grid. Otherwise, consider the microgrid-based distribution transformer asset management solution as optimal.

The optimality of the Benders decomposition method is extensively discussed in [36–38]. A comprehensive discussion on branch-and-bound technique for solving the microgrid-based distribution transformer asset management model is provided in the Appendix.

### 3.2 Microgrid optimal scheduling (master problem)

The objective of the microgrid optimal scheduling master problem is to minimise the microgrid annual operation cost, subject to (13)–(37). The second term added to the objective function is the projected cost of the distribution transformer loss of life, which will be obtained from the optimality cuts generated in the transformer asset management subproblem. The value of this term in the first iteration will be 0. The master problem determines the optimal microgrid schedule, where the optimal values of the exchanged power between the microgrid and the utility grid will be sent to the distribution asset management subproblem with the objective of calculating the optimal value for the distribution transformer loss of life

$$\min \sum_h \sum_t \left[ \sum_{i \in G} F_i(P_{iht}) + \rho_{ht}^M P_{ht}^M \right] + \Lambda \tag{40}$$

s.t. (13)–(37).

### 3.3 Transformer asset management (subproblem)

The objective of the transformer asset management subproblem is to minimise the cost of distribution transformer loss of life based on the IEEE Std. C57.91-2011, as defined in (41), and subject to additional limitations on the distribution transformer loading (38) and (39)

$$\min \quad Q = \sum_h \sum_t \psi f(K_{ht}^I, K_{ht}^U) \tag{41}$$

$$P_{h(t-1)}^M = \hat{P}_{h(t-1)}^M \quad \lambda_{ht} \quad \forall h, \forall t, \tag{42}$$

$$P_{ht}^M = \hat{P}_{ht}^M \quad \mu_{ht} \quad \forall h, \forall t. \tag{43}$$

The exchanged power of the microgrid with the utility grid (transformer loading) is calculated in the master problem and used in the subproblem as given values in (42), (43). $\lambda_{ht}$ and $\mu_{ht}$ are dual variables associated with the initial and ultimate microgrid exchanged power with the utility grid at each time interval, respectively. These dual variables are calculated thorough linearisation of subproblem around the operating point in each iteration, determined in the master problem.

The solution of the original integrated problem based on the current obtained solution would provide an upper bound (44), while the lower bound in each iteration is the solution of the master problem, i.e. microgrid annual operation cost plus the term reflecting the cost of transformer loss of life

$$UB = \sum_h \sum_t \left[ \sum_{i \in G} F_i(\hat{P}_{iht}) + \rho_{ht}^M \hat{P}_{ht}^M \right] + \psi f(\hat{K}_{ht}^I, \hat{K}_{ht}^U) \tag{44}$$

The final solution of the original problem is achieved when the difference between these two bounds is smaller than a threshold. If the convergence criterion is not satisfied, the optimality cut (45), is generated and added to the master problem to revise the solution in the next iteration.

$$\Lambda \geq \hat{Q} + \sum_h \sum_t \lambda_{ht} \left( |P_{h(t-1)}^M| - |\hat{P}_{h(t-1)}^M| \right) + \sum_h \sum_t \mu_{ht} \left( |P_{ht}^M| - |\hat{P}_{ht}^M| \right) \tag{45}$$

$\hat{Q}$ is the calculated objective value for the distribution transformer loss of life (optimal solution for the subproblem). Moreover, the optimality cut (45) consists of two terms associated with the initial and ultimate microgrid exchanged power with the utility grid. This cut indicates that the solution of the revised microgrid optimal scheduling could lead to a better solution for the transformer asset management subproblem, i.e. the one which causes a smaller cost for the distribution transformer loss of life. The absolute function in (45) makes the master problem non-linear. In order to have a linear model in the master problem, two new non-negative variables ($P^{M1}$ and $P^{M2}$) are considered in a way that only one of them can be selected via binary variables $x$ and $y$ (46) and (47). As $P^{M1}$, $P^{M2}$ are both non-negative variables and only one of them can be non-zero at every hour, in case of power export ($P^M>0$) $P^M = P^{M1}$ and $P^{M2} = 0$, and similarly, in case of power import ($P^M<0$) $P^M = -P^{M2}$ and $P^{M1} = 0$

$$P_{ht}^M = x_{ht} P_{ht}^{M1} - y_{ht} P_{ht}^{M2} \quad \forall h, \forall t, \tag{46}$$

$$x_{ht} + y_{ht} \leq 1 \quad \forall h, \forall t. \tag{47}$$

Multiplication of binary variables ($x$ and $y$) with continues variables ($P^{M1}$ and $P^{M2}$) makes bilinear terms ($x_{ht} P^{M1}$ and $y_{ht} P^{M2}$) in (46), which are linearised via (48)–(50), with $M$ as a large positive constant

$$-M x_{ht} - M y_{ht} \leq P_{ht}^M \leq M x_{ht} + M y_{ht} \quad \forall h, \forall t, \tag{48}$$

$$P_{ht}^{M1} - M(1 - x_{ht}) \leq P_{ht}^M \leq P_{ht}^{M1} + M(1 - x_{ht}) \quad \forall h, \forall t, \tag{49}$$

$$-P_{ht}^{M2} - M(1 - y_{ht}) \leq P_{ht}^M \leq -P_{ht}^{M2} + M(1 - y_{ht}) \quad \forall h, \forall t. \tag{50}$$

If binary variables $x$ and $y$ are zero, $P^M$ would be 0 and (49), (50) would be relaxed. If binary variables $x$ or $y$ are 1, (48) would be relaxed and $P^M$ would be equal to either $P^{M1}$ or $-P^{M2}$, based on (49) and (50), respectively. In order to have a positive value for the $P^M$ in (45), this variable is replaced with the summation of $P^{M1}$ and $P^{M2}$ which leads to a revised representation of the optimality cut

$$\Lambda \geq \hat{Q} + \sum_h \sum_t \lambda_{ht} [(P_{h(t-1)}^{M1} + P_{h(t-1)}^{M2}) - (\hat{P}_{h(t-1)}^{M1} + \hat{P}_{h(t-1)}^{M2})] + \sum_h \sum_t \mu_{ht} [(P_{ht}^{M1} + P_{ht}^{M2}) - (\hat{P}_{ht}^{M1} + \hat{P}_{ht}^{M2})] \tag{51}$$

The optimality cut (45) plays a key role in restricting the lower bound of the microgrid optimal scheduling master problem. Using the proposed Benders decomposition procedure in an iterative manner between the master problem and the subproblem, a decomposed model for the microgrid-based distribution transformer asset management will be achieved. This model reaps the benefits of reshaping microgrid exchanged power with the utility grid to maximise the distribution transformer lifetime.

## 4 Numerical simulations

To investigate the performance of the proposed model, a test microgrid which consists of four dispatchable DGs, two non-

**Table 1** Characteristics of generation units

| Unit | Type | Cost coefficient, $/MWh | Min–max capacity, MW | Min up/down time, h | Ramp up/down rate, MW/h |
|---|---|---|---|---|---|
| G1 | D | 27.7 | 1–5 | 3 | 2.5 |
| G2 | D | 39.1 | 1–5 | 3 | 2.5 |
| G3 | D | 61.3 | 0.8–3 | 1 | 3 |
| G4 | D | 65.6 | 0.8–3 | 1 | 3 |
| G5 | ND | 0 | 0–1 | — | — |
| G6 | ND | 0 | 0–1.5 | — | — |

D: dispatchable, ND: non-dispatchable.

**Table 2** Characteristics of the energy storage system

| Storage | Capacity, MWh | Min–max charging/discharging power, MW | Min charging/discharging time, h |
|---|---|---|---|
| ESS | 10 | 0.4–2 | 5 |

**Table 3** Characteristics of adjustable loads

| Load | Type | Min–max capacity, MW | Required energy, MWh | Initial start-end time, h | Min up time, h |
|---|---|---|---|---|---|
| L1 | S | 0–0.4 | 1.6 | 11–15 | 1 |
| L2 | S | 0–0.4 | 1.6 | 15–19 | 1 |
| L3 | S | 0.02–0.8 | 2.4 | 16–18 | 1 |
| L4 | S | 0.02–0.8 | 2.4 | 14–22 | 1 |
| L5 | C | 1.8–2 | 47 | 1–24 | 24 |

S: shiftable, C: curtailable.

dispatchable DGs (G5: wind and G6: solar), one DES, and five adjustable loads is considered and studied. The characteristics of generation units, energy storage system and adjustable loads are tabulated in Tables 1–3, respectively. The forecasted values for microgrid hourly fixed load, non-dispatchable units' generation, and the market price for one sample day are provided in Tables 4–6, respectively. Note that scheduling horizon of one year is considered in this paper. More details on the hourly loads and market price for the considered one-year operation are available in [39]. A 10 MVA distribution transformer is considered at the Point of Common Coupling with the characteristics borrowed from [7]. The nominal active power of the distribution transformer is considered to be 10 MW. In order to calculate the transformer loss of life, the hourly forecasted ambient temperature of a specific location in Houston, TX [40] for one year is used. Since this study does not take into account power congestion and power flow calculations, the system topology diagram is not of significance and the results are independent of the topology.

In order to investigate the effectiveness of the proposed model, the following cases are studied:

*Case 0:* Transformer loss of life calculation.
*Case 1:* Microgrid optimal scheduling ignoring transformer asset management constraints.
*Case 2:* Microgrid optimal scheduling considering transformer asset management constraints.
*Case 3:* Microgrid optimal scheduling with limited transformer overloading while ignoring asset management constraints.
*Case 4:* Microgrid optimal scheduling with limited transformer overloading and asset management constraints.
*Case 5:* Sensitivity analysis with regards to market price forecast errors, transformer loading, and adjustable loads.

*Case 0*: In this case, it is assumed that the microgrid loads are only supplied by the utility grid, i.e. the local generation is ignored. The transformer loading, in this case, is similar to the microgrid load profile, as the exchanged power with the utility grid to supply the microgrid load passes through the transformer. The annual transformer loss of life, in this case, is calculated as 3.1%, which represents an expected lifetime of 32 years.

**Table 4** Microgrid hourly fixed load (one day as a sample)

| time, h | 1 | 2 | 3 | 4 | 5 | 6 |
|---|---|---|---|---|---|---|
| load, MW | 8.73 | 8.54 | 8.47 | 9.03 | 8.79 | 8.81 |
| time, h | 7 | 8 | 9 | 10 | 11 | 12 |
| load, MW | 10.12 | 10.93 | 11.19 | 11.78 | 12.08 | 12.13 |
| time, h | 13 | 14 | 15 | 16 | 17 | 18 |
| load, MW | 13.92 | 15.27 | 15.36 | 15.69 | 16.13 | 16.14 |
| time, h | 19 | 20 | 21 | 22 | 23 | 24 |
| load, MW | 15.56 | 15.51 | 14.00 | 13.03 | 9.82 | 9.45 |

**Table 5** Generation of non-dispatchable units (one day as a sample)

| time, h | 1 | 2 | 3 | 4 | 5 | 6 |
|---|---|---|---|---|---|---|
| G5, MW | 0 | 0 | 0 | 0 | 0.63 | 0.80 |
| G6, MW | 0 | 0 | 0 | 0 | 0 | 0 |
| time, h | 7 | 8 | 9 | 10 | 11 | 12 |
| G5, MW | 0.62 | 0.71 | 0.68 | 0.35 | 0.62 | 0.36 |
| G6, MW | 0 | 0 | 0 | 0 | 0 | 0.75 |
| time, h | 13 | 14 | 15 | 16 | 17 | 18 |
| G5, MW | 0.4 | 0.37 | 0 | 0 | 0.05 | 0.04 |
| G6, MW | 0.81 | 1.20 | 1.23 | 1.28 | 1.00 | 0.78 |
| time, h | 19 | 20 | 21 | 22 | 23 | 24 |
| G5, MW | 0 | 0 | 0.57 | 0.60 | 0 | 0 |
| G6, MW | 0.71 | 0.92 | 0 | 0 | 0 | 0 |

**Table 6** Hourly electricity price (one day as a sample)

| time, h | 1 | 2 | 3 | 4 | 5 | 6 |
|---|---|---|---|---|---|---|
| price, $/MWh | 15.03 | 10.97 | 13.51 | 15.36 | 18.51 | 21.8 |
| time, h | 7 | 8 | 9 | 10 | 11 | 12 |
| price, $/MWh | 17.3 | 22.83 | 21.84 | 27.09 | 37.06 | 68.95 |
| time, h | 13 | 14 | 15 | 16 | 17 | 18 |
| price, $/MWh | 65.79 | 66.57 | 65.44 | 79.79 | 115.45 | 110.28 |
| time, h | 19 | 20 | 21 | 22 | 23 | 24 |
| price, $/MWh | 96.05 | 90.53 | 77.38 | 70.95 | 59.42 | 56.68 |

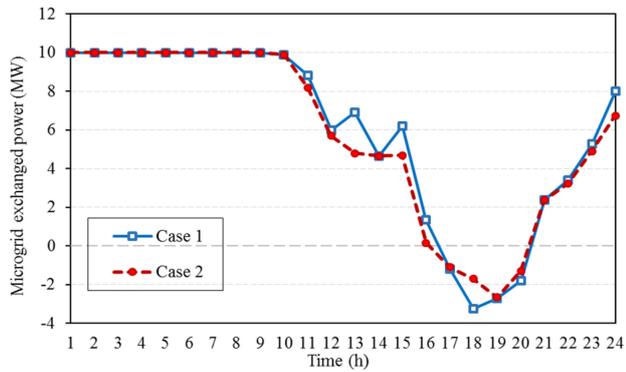

**Fig. 2** *Microgrid exchanged power with the utility grid in Cases 1 and 2 in a sample day of the studied year*

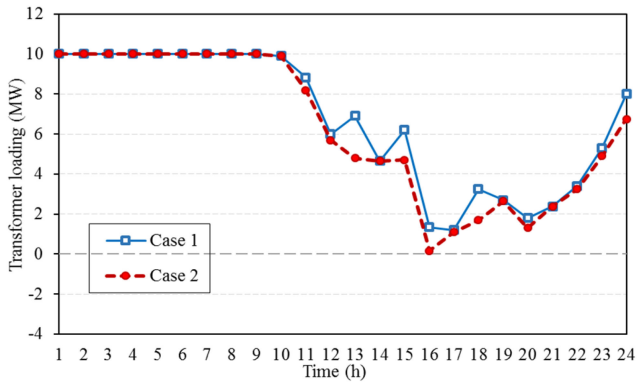

**Fig. 3** *Transformer loading in Cases 1 and 2 in a sample day of the studied year*

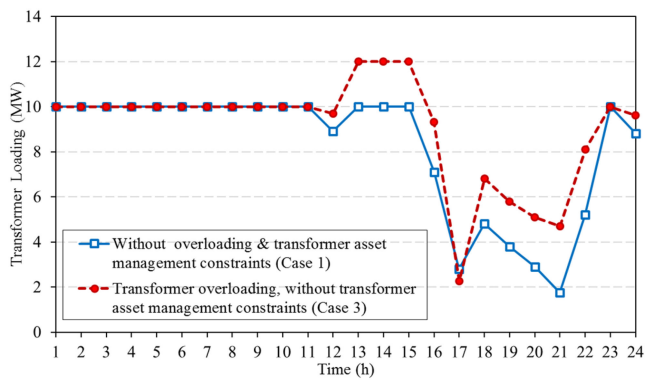

**Fig. 4** *Transformer loading in Case 3 in one of the days with transformer overloading as a sample*

*Case 1*: The grid-connected price-based optimal scheduling is analysed for a one-year horizon. In the price-based scheduling, the main goal is to minimise the microgrid operation cost without any commitments in supporting transformer asset management. The microgrid operation cost is calculated as $1,632,296, and the annual transformer loss of life is calculated as 2.7% in this case. If this value is considered as the average annual loss of life, an expected lifetime of 37 years is perceived for the transformer. The primary reason of this longer lifetime (37 years) compared to the value calculated in Case 0 (32 years) is the microgrid local generation which would partially supply local loads and thus reduce the transformer loading. This situation leads to a smaller loss of life and consequently longer lifetime for the distribution transformer. In other words, even without considering asset management in microgrid scheduling, the transformer lifetime will be prolonged as the microgrid reduces transformer loading through the local generation and partial load offset. It should, however, be noted that possible transformer overloading is ignored in this case.

*Case 2*: In this case, the microgrid controller minimises the microgrid operation cost while considering the transformer asset management constraints. In other words, in addition to minimising the operation cost, the microgrid controller attempts to reduce the transformer loading which leads to lowering the transformer loss of life and consequently translates into a longer lifetime. The annual transformer loss of life is reduced from 2.7% in Case 1 to 2.08%, at the expense of 0.11% increases in microgrid operation cost compared to Case 1 to reach a cost of $1,634,239. The transformer lifetime is increased in this case by an average of 11 years. Two points can be considered here: (i) this considerable increase in the transformer lifetime is achieved by the insignificant addition of less than $2000/year to the microgrid operation cost, and (ii) transformer is not overloaded in any of the operation hours, i.e. the microgrid only reshapes the transformer loading profile without causing any overloads. The considerable impact of overloads will be further discussed in the following cases.

Fig. 2 compares the exchanged power with utility grid in Cases 1 and 2 in one day, as a sample from the one-year optimal scheduling horizon. As the figure shows, as the mere aim of the microgrid in Case 1 is minimising its operation cost, the power is purchased from the utility grid when the market price is low, and the extra power is sold back to the utility grid when the market price is high. In other words, the economic incentive is the only major factor in determining the optimal schedule. However, in Case 2, in addition to microgrid optimal scheduling, the distribution transformer loss of life is considered, so the exchanged power is reshaped in order to reduce load variations. Explicitly power exchange is changed in hours 13, 15, and 18 as it is more economical to reduce the transformer loading rather than purchasing less expensive power from or selling extra power to the utility grid.

Fig. 3 depicts the transformer loading in both cases in the same studied day, which better illustrates the effect of the transformer asset management constraints on the microgrid power exchange. The depicted transformer loading is the absolute value of microgrid exchanged power with the utility grid shown in Fig. 2. As this figure shows the transformer loading is reduced in the range between 0.1 (at hour 17) and 2.1 MW (at hour 13). This decrease causes a reduction in the transformer loss of life in this specific day from 0.0040 to 0.00367%. This reduced rate is the effect of applying transformer asset management in the microgrid optimal scheduling during only one sample day of the studied year.

*Case 3*: The transformer overloading is considered in this case, without taking the transformer asset management constraints into account. A 20% overloading at 3 h (13, 14, and 15) of 20 random days in a year is considered, that is in only 60 h of 8760 h in a year. Fig. 4 shows the transformer loading in this case and compares it with that of Case 1 (without transformer overloading). As Fig. 4 shows, a 3-h overloading in the afternoon not only leads to changes in the transformer loading pattern during the transformer overloaded hours but also impacts the transformer loading in the remaining hours of the studied day. The transformer loss of life, in this case, is increased to 3.09% compared to 2.7% in the case without overloads.

The results show that the initial transformer loss of life of 0.0065% is increased to 0.0264% in this sample day only due to a 3-h overload. This significant rise of the transformer loss of life (more than four times) shows the considerable effect of the transformer overloading on its lifetime reduction. This increase occurs due to the exponential nature of the equations used in calculating the transformer loss of life. The microgrid operation cost, in this case, is calculated as $1,628,345. It should be noted that the sample day, in this case, shown in Figs. 4 and 5, is selected from the twenty studied days for transformer overloading, and it is not the same as the selected day in Figs. 2 and 3.

*Case 4*: The parameters and conditions of this case are similar to those in Case 3, while the transformer asset management constraints are considered as well. By adding the transformer asset management constraints, as Fig. 5 demonstrates, the transformer loading decreases not only during the overloading hours but also at the most hours after the overloading. The changes in microgrid schedule and energy arbitrage lead to a 22% decrease in the transformer loss of life (2.41% in this case compared with 3.09% in Case 3). However, this drop in the transformer loss of life and increasing its lifetime leads to a higher microgrid operation cost, calculated as $1,630,842 in this case.

The obtained results of the studied cases are tabulated in Table 7 As the results of Cases 0 and 1 demonstrate, utilising a microgrid significantly decreases the annual transformer loss of life and consequently increases the expected lifetime of the transformer. A comparison between Cases 1 and 2 advocates that taking transformer asset management constraints into account leads to decreasing the annual transformer loss of life even further (48 years in Case 2, compared to 37 years in Case 1), while the annual microgrid operation cost marginally increases. A comparison between Cases 3 and 4 also highlights the impact of the transformer asset management constraints on reducing the transformer loss of life under transformer overloading conditions.

*Case 5*: The sensitivity of the provided results with regards to market price forecast errors, transformer loading, and adjustable loads are thoroughly investigated in this case.

*Case 5a*: Sensitivity analysis with regards to market price forecast errors: A sensitivity analysis is performed to study the impact of forecast errors on annual transformer loss of life, transformer expected lifetime, and annual microgrid operation cost. Forecast errors of ±10, ±20, and ±30% are considered for the annual hourly market price. The obtained results for this sensitivity analysis are tabulated in Table 8. As the obtained results show, the annual transformer loss of life drops by increasing market price forecast errors, and accordingly the transformer expected lifetime increases. When market price increases, the master controller readjusts the microgrid schedule with the objective of supplying the loads locally rather than importing power from the utility grid. Nevertheless, the microgrid exchanged power with the utility grid, i.e. transformer loading is decreased, which translates into the lower transformer loss of life and a higher transformer expected lifetime, in cases of ignoring and considering transformer asset management constraints. In addition, the results demonstrate that the annual transformer loss of life as well as the transformer expected lifetime, are significantly improved by taking the transformer asset management constraints into account. For instance, in the case of '30% decrease' and '30% increase', the transformer expected lifetime grows 6 and 12.5 years, respectively.

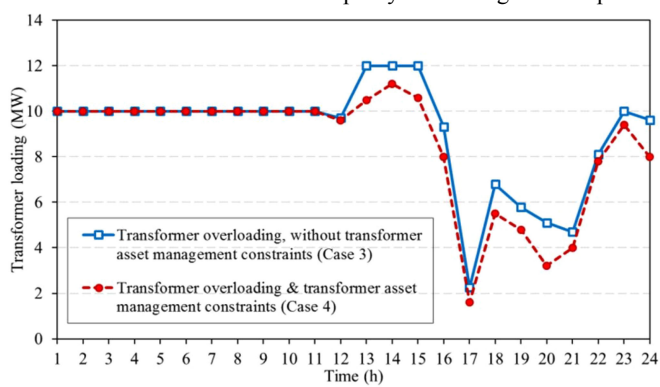

**Fig. 5** *Comparison of transformer loading in Cases 3 and 4, transformer overloading with and without transformer asset management*

**Table 7** Microgrid operation cost and transformer loss of life and lifetime for studied cases

| Case | Annual microgrid operation cost, $ | Annual transformer loss of life, % | Transformer expected lifetime, years |
|---|---|---|---|
| 0 | — | 3.1 | 32 |
| 1 | 1,632,296 | 2.7 | 37 |
| 2 | 1,634,239 | 2.08 | 48 |
| 3 | 1,628,345 | 3.09 | 32.3 |
| 4 | 1,630,842 | 2.41 | 41.5 |

It should be noted that the annual microgrid operation cost slightly raises by considering transformer asset management constraints, in the expense of lowering the transformer loss of life and increasing the transformer expected lifetime.

*Case 5b*: Sensitivity analysis with regards to transformer loading: The effect of transformer loading on the annual transformer loss of life as well as the transformer expected lifetime is investigated in this case. To this end, 50, 75, 100, and 125% of the transformer nominal power ($P_{nom}$) are considered as the maximum limitation for the transformer loading. The obtained results for this study are listed in Table 9. The sensitivity results clearly depict the exponential growth of the transformer loss of life by increasing the transformer loading. By keeping the transformer loading within the limit of 50%, the annual transformer loss of life is calculated, respectively, as 0.455 and 0.452% in cases of ignoring and considering transformer asset management constraints. On the other hand, overloading the distribution transformer will dramatically reduce its lifetime. The transformer loss of life under 125% transformer loading, i.e. 25% overload, is, respectively, calculated as 11.83 and 8.61% in cases of ignoring and considering transformer asset management constraints, where accordingly the transformer expected lifetime will be 8.5 and 11.6 years, respectively. Moreover, the results demonstrate that the transformer expected lifetime will be increased slightly while taking the transformer asset management constraints into account for lower transformer loading limits. It should be noted that the cases with very low/high limits of the transformer loading, i.e. 50 or 125%, are not practical and just are considered in this study as extreme operating conditions.

*Case 5c*: Sensitivity analysis with regards to adjustable loads: To demonstrate the effect of adjustable loads on the annual transformer loss of life, transformer expected lifetime, and annual microgrid operation cost, the problem is solved for various cases of adjustable loads. The required energy of the five aggregated adjustable loads is changed from 10 to 50 MWh (which however can be considered as having more adjustable loads in the microgrid). The obtained results for this study are provided in Table 10. As the sensitivity analysis results show, by increasing the adjustable loads, the annual transformer loss of life slightly lessens, which means the transformer expected lifetime increases. By changing the total required energy of adjustable loads from 10 to

**Table 8** Sensitivity analysis with regards to market price forecast error

| Market price | Annual transformer loss of life, % | | Transformer expected lifetime, years | | Annual microgrid operation cost, $ | |
|---|---|---|---|---|---|---|
| | Ignoring transformer asset management constraints | Considering transformer asset management constraints | Ignoring transformer asset management constraints | Considering transformer asset management constraints | Ignoring transformer asset management constraints | Considering transformer asset management constraints |
| 30% decrease | 3.41 | 2.83 | 29.3 | 35.3 | 1,242,627 | 1,245,215 |
| 20% decrease | 3.077 | 2.41 | 32.5 | 41.5 | 1,396,111 | 1,399,733 |
| 10% decrease | 2.84 | 2.23 | 35.2 | 44.8 | 1,525,675 | 1,528,842 |
| default | 2.7 | 2.08 | 37.0 | 48.1 | 1,632,296 | 1,634,239 |
| 10% increase | 2.57 | 2.011 | 38.9 | 49.7 | 1,715,356 | 1,717,944 |
| 20% increase | 2.51 | 1.935 | 39.8 | 51.7 | 1,776,963 | 1,779,887 |
| 30% increase | 2.456 | 1.88 | 40.7 | 53.2 | 1,821,077 | 1,823,412 |

**Table 9** Sensitivity analysis with regards to transformer loading

| Transformer loading, % | Annual transformer loss of life, % | | Transformer expected lifetime, years | |
|---|---|---|---|---|
| | Ignoring transformer asset management constraints | Considering transformer asset management constraints | Ignoring transformer asset management constraints | Considering transformer asset management constraints |
| 50 | 0.455 | 0.452 | 219.8 | 221.2 |
| 75 | 1.67 | 1.38 | 59.9 | 72.5 |
| 100 | 2.7 | 2.08 | 37 | 48.1 |
| 125 | 11.83 | 8.61 | 8.5 | 11.6 |

**Table 10** Sensitivity analysis with regards to adjustable load

| Adjustable load | Annual transformer loss of life, % | | Transformer expected lifetime, years | | Annual microgrid operation cost, $ | |
|---|---|---|---|---|---|---|
| | Ignoring transformer asset management constraints | Considering transformer asset management constraints | Ignoring transformer asset management constraints | Considering transformer asset management constraints | Ignoring transformer asset management constraints | Considering transformer asset management constraints |
| default | 2.704 | 2.081 | 36.98 | 48.05 | 1,632,296 | 1,634,239 |
| 10 MWh increase | 2.698 | 2.071 | 37.07 | 48.30 | 1,590,523 | 1,590,890 |
| 20 MWh increase | 2.684 | 2.060 | 37.25 | 48.55 | 1,553,959 | 1,557,723 |
| 30 MWh increase | 2.672 | 2.050 | 37.43 | 48.78 | 1,520,413 | 1,524,829 |
| 40 MWh increase | 2.663 | 2.041 | 37.56 | 49.00 | 1,496,362 | 1,499,589 |
| 50 MWh increase | 2.650 | 2.030 | 37.73 | 49.25 | 1,476,587 | 1,478,510 |

50 MWh, the transformer expected lifetime increases by 1.2 years from 48.05 to 49.25 years, when taking the transformer asset management constraints into account. In addition, as the total required energy of adjustable loads increase, the annual microgrid operation cost reduces in both cases of ignoring and considering transformer asset management constraints. Nevertheless, adjustable loads play a key role in reshaping the loading of the distribution transformer at the point of interconnection in order to increase its lifetime. The cost associated with the power loss is extremely smaller than the transformer loss of life and microgrid operation costs so that its impacts will be negligible. Nevertheless, in order to ensure this assumption, a case study is performed in which 6% distribution power loss is considered in the distribution deployed microgrid. The obtained results demonstrate that cost associated with the power loss is a very small fraction of the transformer loss of life and microgrid operation costs. Thus, if the power loss cost of the microgrid is taken into consideration, the results will be affected to a minimal extent; however, the final assessment and conclusion remain intact.

## 5 Conclusion

In this paper, a microgrid-based distribution transformer asset management model was proposed and formulated. Using a Benders decomposition method, the proposed model was decomposed into a microgrid optimal scheduling master problem and a distribution transformer asset management subproblem. Based on a relevant IEEE Standard, the optimal cost of the distribution transformer loss of life was calculated in the subproblem in order to examine the optimality of the microgrid scheduling solution. This means that the distribution transformer asset management subproblem was presented to manipulate the distribution transformer loading via scheduling microgrid resources in an efficient and asset management-aware manner. Numerical simulations were carried out for various conditions of transformer loading to show the advantages and the effectiveness of the proposed model. The results showed that the utility companies can efficiently manage their resources to decrease the transformer loss of life and consequently ensure a considerable increase in transformer lifetime.

## 6 References

[1] Zhang, X., Gockenbach, E.: 'Asset-management of transformers based on condition monitoring and standard diagnosis', *IEEE Electr. Insul. Mag.*, 2008, **24**, (4), pp. 26–40
[2] Abu-Elanien, A.E.B., Salama, M.M.A.: 'Asset management techniques for transformers', *Electr. Power Syst. Res.*, 2010, **80**, (4), pp. 456–464
[3] Ma, H., Saha, T.K., Ekanayake, C., *et al.*: 'Smart transformer for smart grid intelligent framework and techniques for power transformer asset management', *IEEE Trans. Smart Grid*, 2015, **6**, (2), pp. 1026–1034
[4] Abiri-Jahromi, A., Parvania, M., Bouffard, F., *et al.*: 'A two-stage framework for power transformer asset maintenance management – part I: models and formulations', *IEEE Trans. Power Syst.*, 2013, **28**, (2), pp. 1395–1403
[5] Arshad, M., Islam, S.M., Khaliq, A.: 'Power transformer asset management'. Proc. Int. Conf. Power System Technology (PowerCon), November 2004, pp. 1395–1398
[6] Žarković, M., Stojković, Z.: 'Analysis of artificial intelligence expert systems for power transformer condition monitoring and diagnostics', *Electr. Power Syst. Res.*, 2017, **149**, pp. 125–136
[7] Muthanna, K.T., Sarkar, A., Das, K., *et al.*: 'Transformer insulation life assessment', *IEEE Trans. Power Deliv.*, 2006, **21**, (1), pp. 150–156
[8] Hilshey, A., Hines, P., Rezaei, P., *et al.*: 'Estimating the impact of electric vehicle smart charging on distribution transformer aging', *IEEE Trans. Smart Grid*, 2013, **4**, (2), pp. 905–913
[9] Sarfi, V., Mohajeryami, S., Majzoobi, A.: 'Estimation of water content in a power transformer using moisture dynamic measurement of its oil', *IET High Voltage*, 2017, **2**, (1), pp. 11–16
[10] Singh, J., Sood, Y.R., Verma, P.: 'Experimental investigation using accelerated aging factors on dielectric properties of transformer insulating oil', *Electr. Power Compon. Syst.*, 2011, **39**, pp. 1045–1059
[11] Seier, A., Hines, P.D.H., Frolik, J.: 'Data-driven thermal modeling of residential service transformers', *IEEE Trans. Smart Grid*, 2015, **6**, (2), pp. 1019–1025
[12] IEEE Std. C57.91-2011: 'IEEE guide for loading mineral-oil-immersed transformers and step-voltage regulators', 2012
[13] Mahoor, M., Majzoobi, A., Hosseini, Z.S., *et al.*: 'Leveraging sensory data in estimating transformer lifetime'. North American Power Symp. (NAPS), Morgantown, WV, September 2017
[14] Arshad, M., Islam, S.S.: 'A novel fuzzy logic technique for power asset management'. Proc. Industry Applications Conf., October 2006, pp. 276–286
[15] Majzoobi, A., Mahoor, M., Khodaei, A.: 'Machine learning applications in estimating transformer loss of life'. IEEE PES General Meeting, Chicago, IL, July 2017
[16] Mahoor, M., Khodaei, A.: 'Data fusion and machine learning integration for transformer loss of life estimation'. IEEE PES Transmission and Distribution Conf. and Exposition (T&D), Denver, CO, April 2018
[17] Velasquez-Contreras, J.L., Sanz-Bobi, M.A., Galceran Arellano, S.: 'General asset management model in the context of an electric utility: application to power transformers', *Electr. Power Syst. Res.*, 2011, **81**, (11), pp. 2015–2037
[18] Aravinthan, V., Jewell, W.: 'Controlled electric vehicle charging for mitigating impacts on distribution assets', *IEEE Trans. Smart Grid*, 2015, **6**, (2), pp. 999–1009
[19] Roe, C., Evangelos, F., Meisel, J., *et al.*: 'Power system level impacts of PHEVs'. Proc. 42nd Hawaii Int. Conf. Systems Science, January 2009, pp. 1–10
[20] Gray, M.K., Morsi, W.G.: 'On the impact of single-phase plug-in electric vehicles charging and rooftop solar photovoltaic on distribution transformer aging', *Electr. Power Syst. Res.*, 2017, **148**, pp. 202–209
[21] Gray, M.K., Morsi, W.G.: 'On the role of prosumers owning rooftop solar photovoltaic in reducing the impact on transformer's aging due to plug-in electric vehicles charging', *Electr. Power Syst. Res.*, 2017, **143**, pp. 563–572

[22] Shokrzadeh, S., Ribberink, H., Rishmawi, I., *et al.*: 'A simplified control algorithm for utilities to utilize plug-in electric vehicles to reduce distribution transformer overloading', *Energy*, 2017, **133**, pp. 1121–1131
[23] Ramos Munoz, E., Razeghi, G., Zhang, L., *et al.*: 'Electric vehicle charging algorithms for coordination of the grid and distribution transformer levels', *Energy*, 2016, **113**, pp. 930–942
[24] 'Department of Energy Office of Electricity Delivery and Energy Reliability Smart Grid R&D Program-Summary Report: 2012 DOE Microgrid Workshop'. Available at https://www.energy.gov/sites/prod/files/2012%20Microgrid%20Workshop%20Report%2009102012.pdf, accessed 02 April 2018
[25] Majzoobi, A., Khodaei, A.: 'Application of microgrids in supporting grid flexibility', *IEEE Trans. Power Syst.*, 2017, **32**, (5), pp. 3660–3669
[26] Shi, L., Luo, Y., Tu, G.Y.: 'Bidding strategy of microgrid with consideration of uncertainty for participating in power market', *Int. J. Electr. Power Energy Syst.*, 2014, **59**, pp. 1–13
[27] Majzoobi, A., Khodaei, A.: 'Application of microgrids in providing ancillary services to the utility grid', *Energy*, 2017, **123**, pp. 555–563
[28] Khodaei, A., Bahramirad, S., Shahidehpour, M.: 'Microgrid planning under uncertainty', *IEEE Trans. Power Syst.*, 2015, **30**, (5), pp. 2417–2425
[29] Khodaei, A.: 'Microgrid optimal scheduling with multi-period islanding constraints', *IEEE Trans. Power Syst.*, 2014, **29**, (3), pp. 1383–1392
[30] Khodaei, A.: 'Resiliency-oriented microgrid optimal scheduling', *IEEE Trans. Power Syst.*, 2014, **5**, (4), pp. 1584–1591
[31] Bahramirad, S., Reder, W., Khodaei, A.: 'Reliability-constrained optimal sizing of energy storage system in a microgrid', *IEEE Trans. Smart Grid, Special Issue on Microgrids*, 2012, **3**, (4), pp. 2056–2062
[32] Shahidehpour, M., Clair, J.: 'A functional microgrid for enhancing reliability, sustainability, and energy efficiency', *Electr. J.*, 2012, **25**, (8), pp. 21–28
[33] 'DOE Microgrid Workshop Report'. Available at https://www.energy.gov/sites/prod/files/Microgrid%20Workshop%20Report%20August%202011.pdf, accessed 02 April 2018
[34] 'Microgrid Deployment Tracker 4Q17'. Available at https://www.navigantresearch.com/research/microgrid-deployment-tracker-4q17, accessed 02 April 2018
[35] Parhizi, S., Lotfi, H., Khodaei, A., *et al.*: 'State of the art in research on microgrids: a review', *IEEE. Access*, 2015, **3**, pp. 890–925
[36] Geoffrion, A.M.: 'Generalized benders decomposition', *J. Optim. Theory Appl.*, 1972, **10**, (4), pp. 237–260
[37] Shahidehpour, M., Fu, Y.: 'Benders decomposition', *IEEE Power Energy Mag.*, 2005, **3**, (2), pp. 20–21
[38] Conejo, A.J., Castillo, E., Minguez, R., *et al.*: '*Decomposition techniques in mathematical programming: engineering and science applications*' (Springer Science & Business Media, New York, NY, USA, 2006)
[39] 'PEAK Lab'. Available at https://portfolio.du.edu/peaklab/page/63877, accessed 02 April 2018
[40] 'Weather history & data archive'. Available at https://www.wunderground.com/history/airport, accessed 02 April 2018
[41] Smith, J.C., Taskin, Z.C.: 'A tutorial guide to mixed-integer programming models and solution techniques', in Lim, G.J., Lee, E.K. (Eds.): '*Optimization in Medicine and Biology*' (Taylor and Francis, Auerbach Publications, New York, NY, USA, 2008), pp. 521–548
[42] Gupta, O.K., Ravindran, A.: 'Branch and bound experiments in convex nonlinear integer programming', *Manage. Sci.*, 1985, **31**, (12), pp. 1533–1546

## 7 Appendix

Branch-and-bound is a commonly-used technique for solving mixed-integer linear programming problems. Two processes are employed in this technique (i) bounding process and (ii) branching process. In the bounding process, the solution of a relaxed mixed-integer linear programming problem, i.e. converting mixed-integer linear programming problem into liner programming problem via removing integrity restrictions, is calculated and then imposed as lower bound for minimisation problems or upper bound for maximisation problems. In the branching process, the problem is broken into two subproblems, where further are solved to obtain the solutions. If the solutions for both of these subproblems satisfy the integrity conditions, they are compared with each other, and the subproblem solution related to smaller objective function value for minimisation problem or a larger one for maximisation problem will be selected as the optimal solution. Note that if only one of these two subproblems solutions satisfies the mixed-integer linear programming integrity condition, this solution is kept as an incumbent solution (i.e. the optimal solution if no better solution will be achieved further). Nevertheless, the branching process is continued to search on the other subproblem with the objective of finding a better solution that satisfied the mixed integer linear programming integrity condition [41].

Mixed-integer linear programming solvers, including but not limited to CPLEX, Xpress-MP, SYMPHONEY, and CBC, reap the benefits of a combination of branch-and-bound techniques and cutting-plane techniques to accelerate the computation time associated with solving mixed-integer linear programming problems, which consequently facilitate solving large mixed-integer linear programming problems using personal computers.

The branch-and-bound technique for solving mixed-integer non-linear programming problems is based on the same idea as the branch-and-bound technique employed to solve mixed-integer linear programming problems. Similar to the branch-and-bound technique explained above, the technique starts by solving the problem in where the discrete conditions of the binary variables are relaxed. If the obtained solution is integral, then this solution is considered as an optimal solution for the problem. Without loss of generality, the two processes of bounding and branching are employed in order to find the optimal solution for the mixed-integer non-linear programming problem [42].